\documentstyle[prl,aps,amssymb,graphicx]{revtex}
\newcommand{\met}{\frac{1}{2}}
\newcommand{\beq}{\begin{equation}}
\newcommand{\eneq}{\end{equation}}

\input{epsf}

\begin{document}

\tolerance 10000

\twocolumn[\hsize\textwidth\columnwidth\hsize\csname %
@twocolumnfalse\endcsname

\draft

\title{Josephson Current in a Quantum Dot in the Kondo Regime Connected to
Two Superconductors}

\author {G. Campagnano$^{*}$, D. Giuliano$^\dagger$ and A. Tagliacozzo$^{*, 
\dagger}$}

\address{$^*$ Dipartimento di Scienze Fisiche, 
        Universit\`a degli Studi di Napoli ``Federico II\\
        $^\dagger$ I.N.F.M., Unit\`a di Napoli, Monte S.Angelo - via Cintia, 
        I-80126 Napoli, Italy}

\date{\today}
\maketitle
\widetext

\begin{abstract}
\begin{center}

\parbox{14cm}{We apply a Gutzwiller-like variational technique to study 
Josephson conduction across a quantum dot with an odd number of electrons
connected to two superconducting leads. We show that, for small values of
the superconducting gap, Kondo correlations and superconductivity cooperate
to enhance the Josephson current. As the superconducting gap increases, 
the current changes sign and the system becomes a $\pi$-junction. The 
$\pi$-junction behavior sets in much before antiferromagnetic correlations at
the dot can be treated perturbatively.}

\end{center}
\end{abstract}

\pacs{
\hspace{1.9cm}
}
]

\narrowtext
Phase coherent transport in  hybrid  superconducting-semiconducting 
nanostructures is already extensively investigated. In these devices,  
quite interesting and surprising features  emerge,   due to electron-electron 
interaction \cite{raimondi}.  In this letter we study a Quantum Dot (QD) at 
Coulomb Blockade (CB) with an odd number of electrons $N$, connected to 
superconducting contacts \cite{zaikin}. Although such a kind of system has 
not been realized yet, it is very likely that continuous improvement of  
nanofabrication techniques will soon make it available. 
In particular, we address the question whether the formation of a Kondo 
singlet at the dot competes or cooperates with BCS s-wave pairing in the 
leads. We show  that, as long as $\Delta \ll k_B T_K$, where $T_K$ is the 
Kondo temperature of the dot and $\Delta$ is the superconducting gap, 
superconductivity does not compete with Kondo 
correlation, but rather favors it. The system behaves as a regular
Josephson junction and the critical Josephson current $I_J$ is even enhanced 
by Kondo  coupling. However, as $\Delta$ becomes comparable with $k_B T_K$, 
there is a crossover to a regime  where superconductivity 
spoils  Kondo  correlations on the dot, characterized by $\pi$-Josephson 
coupling across the dot \cite{clerk}.

Coulomb Blockade  has been widely investigated in Quantum Dots with 
normal leads \cite{kouwenhoven}. If the coupling with the contacts is weak 
and the charging energy is higher than the thermal activation energy, 
DC conduction is strongly  dependent on  the gate voltage $V_g$. The  DC 
conductance across the dot  shows a sequence of peaks, corresponding to 
resonances between the chemical potential of the leads $\mu$  and an energy 
level of the dot. As the dot is tuned in a ``Coulomb Blockade'' (CB) valley 
between two consecutive peaks, $N$ is fixed and the conductance is heavily 
suppressed.

In  Superconducting/Normal/Superconducting (SNS) structures, Cooper pair
subgap tunneling current takes place via Andreev states localized in the 
normal region \cite{andreev}. 
When the normal region is given by a QD, charge quantization at the dot is not
spoiled as the contacts become superconducting \cite{matgla}. The $I-V$
curve of a dot at CB between two superconductors has been
derived in \cite{zaikin}, showing a rather interesting interplay between
multiple Andreev reflection and electronic interaction at the QD.  

Kondo conductance may be achieved in a QD with $N$ odd within a CB valley 
by increasing the coupling between dot and leads, provided  the temperature 
$T$ $<T_K$ \cite{goldhaber}. At $T < T_K$, a strongly correlated state 
between dot and leads sets in and a resonance in the density of states of
the QD opens up at $\mu$. Correspondingly, DC conductance across the dot 
increases, until it eventually reaches the 
unitarity limit at $T=0$\cite{nozieres,silvano}.
Since charge dynamics is ``frozen out'' by CB, the QD in this regime is 
usually modelized as a magnetic impurity with total spin  $S= \met$. 

Magnetic impurities embedded in a bulk superconductor are known to strongly
influence the superconducting  critical  temperature. Adding impurity 
states at energies below  the superconducting gap can even give raise to
gapless superconductivity \cite{abrikosov}. 

Tunneling across a magnetic impurity between superconductors with on site 
Coulomb repulsion has been recently revisited \cite{matveev,spivak,clerk}. 
If   $\Delta \geq k_B T_K$, the system is unable to scale toward the 
strongly-coupled Kondo regime. 
 Sub-gap Cooper pair  is strongly suppressed by Coulomb repulsion,
unless each co-tunneling step is accompanied by a spin flip at the impurity.
It has been proposed that such a  mechanism may reverse the sign of the 
Josephson current through the dot so that the Josephson energy is at a 
minimum for $\varphi = \pi $, where $\varphi$ is the phase difference 
between the order parameters of the two superconductors attached 
to the impurity  ($\pi$-junction) \cite{spivak}.

We expect that fine tuning of the parameters of a S-QD-S device
(for instance, one may drive the system to a level crossing by tuning 
an external applied magnetic field which,
under appropriate  conditions, does not affect Kondo conduction
\cite{noi,nazarov,silvano}) may allow for the  investigation of the full range 
of physical  conditions, from  $\Delta \ll k_B T_K$  when the system can flow 
toward the strongly coupled Kondo regime prior to 
the onset of superconductivity in the leads, to
 $\Delta \gg k_B T_K$,  when  perturbation theory holds.

We study the  $T =0 $ case with $ \Delta \ll k_B T_K$ using  a 
nonperturbative variational 
technique. This technique has already been applied to the case of a 
dot with normal 
contacts, and it has been shown to provide good qualitative results in
the perturbative regime  as well, where it reproduces the  poor man's 
scaling equation \cite{nos}.

To construct the trial state,  we start from the state  
$| \varphi, s \rangle$, given by the  product of the left (L) condensate 
times the right (R)  condensate, with the two order parameters having a 
phase difference   
$\varphi$, 
times the state of the dot: 
$|s \rangle$.  
\beq
| \varphi , s \rangle = | {\rm BCS}, \; L \rangle \times 
| {\rm BCS}, \; R, \; \varphi \rangle \times | s \rangle
\; .
\eneq
\noindent 
Here  $ s$ is the spin component  along the quantization axis  of the 
dot spin ${\bf S}_d = \met $, at CB with odd $N$.

The minimal model for the Kondo interaction between electrons localized 
on the QD  and elctrons from  the contacts 
 is $ H_K = J \vec{\sigma} (0)\cdot \vec{S}_d $.
 The spin density operator of the delocalized electrons, 
$\vec{\sigma} ( 0 )$, at 
the position of the dot  $x=0$ along the  vertical axis, is :
\[
\vec{\sigma} ( 0 ) = \frac{1}{2 {\cal {V}}} 
\sum_{q , q^{'}} (c_{L,q,\sigma}^\dagger + 
c_{R , q , \sigma}^\dagger ) \vec{\tau}_{\sigma , \sigma^{'}} 
(c_{L,q^{'} ,\sigma^{'} } +  c_{R , q^{'}  , \sigma^{'} } )
\]
\noindent 
We have used  the fermion operators $c_{q,j}(c ^\dagger_{q,j})
 \:(j= L,R) $  in the
plane wave representation to describe the contacts particles and 
${\cal {V}}$ is the normalization volume of the leads.
We take the symmetric case in which hybridization of the dot with the 
$L$ and $R$ contacts, $\Gamma$, is the same. 
  Kondo coupling is antiferromagnetic (AF): $J > 0 $. The interaction term $H_K$ can be obtained from  a Schrieffer Wolff 
transformation  as in the case of normal contacts.
 The  superconductivity  in the contacts does not affect this derivation,
provided $D \gg \Delta$, where $D$ is the bandwidth of the itinerant 
electrons.  

The total Hamiltonian is $H = 
H_S + H_K $, where  $H_S $,defined in eq.(\ref{bcsh}) below.
is the Hamiltonian for the $L$ and $R$ 
superconducting contacts in the $BCS $ approximation.  

 The  correlated  trial state 
is constructed by applying  a Gutzwiller-like projector $P_g$ 
to $ | \varphi , s  \rangle$. As $T \rightarrow 0$, the system scales towards 
the strongly coupled -large $J$- regime. Correspondingly, 
$P_g$ gradually projects out the high-energy 
components of the trial state, 
so that eventually  only a localized spin singlet  survives at the QD.

The  ``projector'' $P_g$ is defined as\cite{nos}:
\[
P_g = \left(1 - \frac{3}{4} g \right) + g ( \vec{\sigma} ( 0 ))^2 -
4 {\bf S}_d \cdot \vec{\sigma } ( 0 ) 
\]
\noindent
$g$ is a  variational parameter, which ranges between $g=0$ and $g=4/3$.
When  $g=0$, we have $P_0 = 1 - 4 {\bf S}_d \cdot \vec{\sigma } ( 0 )$.
$P_0$  
fully projects out the high energy localized spin triplet at $x=0$. As
$g$ varies from $0$ to $4/3$, also the localized spin doublet state it 
increasingly projected
out.  Eventually, when   $g$  reaches the value $g = 4/3$ , 
 only the localized spin singlet is left over.

The trial state is defined as:

\beq
| g , \varphi \rangle = P_g | \varphi , s \rangle
\label{eq2}
\eneq
\noindent
 The  value of $g (J) $  is determined
by finding the minimum of the energy functional $ E [ g , J , \varphi , 
\Delta]$, defined as:

\beq
E [ g , J , \varphi , \Delta ] = \frac{ \langle g , \varphi | H | g , 
\varphi \rangle}{ \langle g , \varphi | g , \varphi \rangle }
\label{eq4}
\eneq
\noindent
Eq.(\ref{eq4})  can be calculated by first expressing the products 
$P_g H P_g$ and
$ (P_g)^2$ in terms of the usual fermion quasiparticle operators 
$
\alpha_{j, q} (\alpha^\dagger _{j, q}),
\beta_{j , q}
(\beta ^\dagger _{j , q}) \:\:(j= L,R )$  which destroy  
(create) excitations on the BCS  states  of the $L$ and $R $ contact and 
then by normal-ordering the corresponding operator products. The operators are:
\begin{eqnarray}
\alpha_{j, q} = u_q c_{j,q,\uparrow} - v_q e^{ i \phi_j} 
c_{j,-q,\downarrow}^\dagger
\nonumber\\
\beta_{j , -q} = 
u_q c_{j , -q , \downarrow} + v_q e^{i \phi_j} 
c_{j, q , \uparrow}^\dagger .
\label{eq1}
\end{eqnarray}
$u_q$ and $v_q$ are the  BCS coherence factors and  
$\phi_R = \varphi$ while  $\phi_L=0$.
 The Hamiltonian for the contacts $ H_S$ is conveniently expressed 
in terms of these  operators as:
\beq
H _S = E_{\rm BCS} + \sum_{q , j = L, R} E_q ( \alpha_{j,q}^\dagger 
\alpha_{j,q} + \beta_{j,-q}^\dagger \beta_{j,-q} )\: . 
\label{bcsh}
\eneq
Here 
 $E_{\rm BCS}$ is the 
total ground state energy of the condensates and  
$E_q = \sqrt{\xi_q^2 + \Delta^2}$ are the energies of the quasiparticle 
excitations  with   
$\xi_q = q^2 / 2 m - \mu$, (we put $\hbar= k_B = 1$ throughout the paper).

The variation of the energy due to the AF coupling at the quantum dot in units
of the bandwidth $D$, $ \epsilon [ \xi , j  , \delta , \varphi ] 
\equiv  E [ g , J  , \Delta , \varphi ]/D $,
takes a simple form once expressed in terms of the
parameters $\xi =  \frac{1}{2} \left( 1- \frac{3}{4} g \right)$,
 $j = 3J / 4D $ and  $\delta = \Delta / D$:
\[
 \epsilon [ \xi , j  , \delta , \varphi ]   =
-  2 j \frac{ 1 - \frac{1}{2} \delta^2 ( 1 + \cos ( \varphi))/
N^2 ( 0 ) \lambda^2 }{ ( 1 + \xi^2)
+ ( \xi^2 - 1 ) \delta ( 1 + \cos ( \varphi ) )/ ( 2 N ( 0 ) \lambda ) } 
\]
\beq
+ \frac{ 2 ( 1 - \xi + \xi^2) \left[ \sqrt{ 1 + \delta^2} - \delta^2 ( 1 + 
\cos ( \varphi ) )/  N ( 0 ) \lambda \right] }{  ( 1 + \xi^2)
+ ( \xi^2 - 1 ) \delta ( 1 + \cos ( \varphi ) )/ ( 2 N ( 0 ) \lambda )} .
\label{eq5}
\eneq
\noindent

Here $\lambda$ is the BCS electron-electron interaction strength and 
 $N (0)$ is the normal phase density of states at the Fermi level for
each spin polarization. The first term is the expectation value of $H_K$,  
while the second is the raise in the average value of the kinetic energy  
Hamiltonian, $ H_S$,  due to the formation of the   singlet  between 
the QD and the environment. The value of  $\xi_{\rm min} $, at which 
the energy is at a minimum measures how much higher-energy 
spin states are projected out: $\xi_{\rm min} =0 $  corresponds to full 
projection of states different from a localized spin singlet at the impurity.

At $\delta = 0 $ the strong coupling fixed point is  $j \to \infty $  and 
$ \xi _{\rm min} (j) \to 0$. In correspondence of the fixed point, there is 
a large  decrease of the Kondo energy:
\beq
\Omega _K = 2 \frac{1 - \xi_{\rm min} + ( \xi_{\rm min} )^2 - j}{ 1 + 
( \xi_{\rm min} )^2} \: .
\label{omk}
\eneq

\begin{figure}
\centering \includegraphics*[width=0.9\linewidth]{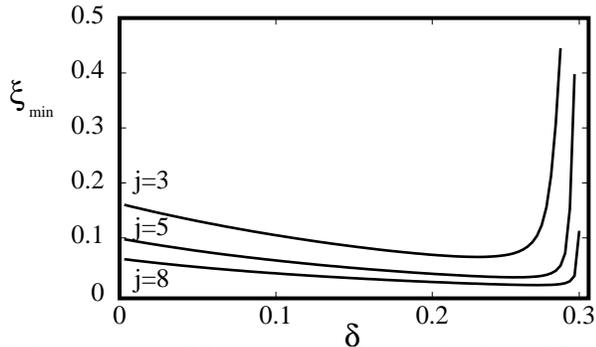}
\caption{$\xi_{\rm min}$ vs. $\delta$ for various $j$. For small values of 
$\delta$, $\xi_{\rm min}$ is lowered, showing that a small amount of 
superconductivity favors Kondo correlations. On the other hand, for
large $\delta$, $\xi$ abruptly increases, that is a signal of the loss
of Kondo regime.}
\label{figura1}
\end{figure}

In Fig.\ref{figura1} we plot $\xi_{\rm min} $ vs.  $\delta $  for various $j$.
As $\delta $ is increased  at $j$ large, $\xi_{\rm min}$ moves from 0 to 
finite values.

As $\delta$ increases, two  regimes can be  envisaged:

{\sl $a)  \:\:\Delta  \ll  T_K $  :
The Kondo correlated state is firmly established.}

The change of  $\xi_{\rm min}$ to first order in $\delta$ is given by:

\[
\xi_{\rm min} ( \delta) = \xi_{\rm min} ( \delta = 0) - \frac{ \delta
 ( 1 + \cos \varphi )}{ N(0) 
\lambda} \left[ \frac{ j+1 -\sqrt{ j^2 + 1} }{ 
 \sqrt{ j^2 + 1}} \right]
\]
\noindent
that shows that, for any value of $\delta$, $\xi_{\rm min} ( \delta ) 
\leq \xi_{\rm min} ( \delta = 0 )$. We also see that, as $\varphi$
changes from 0 to $\pi$, $\xi_{\rm min}$ increases. This clearly 
shows that the stable configuration for the system is at 
$\varphi =0$, what  has important consequences for the Josephson conduction, 
as we are going to show in the following.

\begin{figure}
\centering \includegraphics*[width=0.9\linewidth]{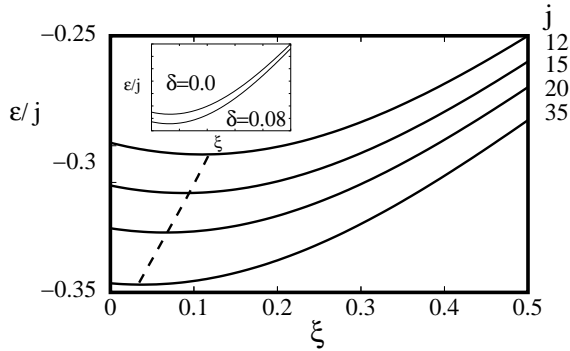}
\caption{$\epsilon/j$ vs. $\xi$ for different $j$ and for $\varphi=0$. At
the onset of Kondo regime (large $j$), the minimum of the curve moves 
toward $\xi=0$. Inset: $\epsilon /j$ vs. $\xi$ for two
different values of $\delta$. A small superconducting gap appears to lower
$\epsilon/j$, that is, a small $\delta$ favors Kondo regime.}
\label{figura2}
\end{figure}

In Fig.\ref{figura2} we plot $ \epsilon  /j $ vs. $\xi$ for various 
values of $j$ at  $\delta=0.03$ and $\varphi=0$. As $j$ increases, 
$\xi_{\rm min}$ moves toward $\xi_{\rm min}=0$, which corresponds to the
strongly coupled fixed point. From the inset in Fig.\ref{figura2} we also 
see that, as $\delta$ increases, the value of the energy at the minimum 
decreases. In this regime  $\epsilon^* [ j ,\delta, \varphi]   <  
\epsilon^* [ j,0, \varphi ] $ that 
is,  superconductivity favors strong Kondo correlations.

Upon calculating $\epsilon [ \xi , j , \delta, \varphi ]$ for 
$\xi = \xi_{\rm min}$,
we get the best estimate for $\epsilon ^* [ j ,\delta ,
\varphi ]$, the energy of the correlated state, to first order in $\delta$:
\beq
\epsilon ^* [ j , \delta , \varphi ] = \Omega _K 
 \left[ 1 +   B \delta 
( 1 + \cos \varphi ) \right] 
\label{eq7}
\eneq
\noindent
where  $B  = [1 - (\xi_{\rm min} )^2]/(  N  (0) \lambda [ 1 +
 (\xi_{\rm min} )^2] )$. From Eq.(\ref{eq7}) we clearly see that, for
large values of $j$, $\epsilon ^* [ j , \delta , \varphi ] <
\epsilon ^* [ j , \delta =0, \varphi  ]$, that is, superconductivity and 
Kondo effect cooperate.

{\sl $b) \:\: \Delta \leq  T_K $ : Superconductivity competes with Kondo 
ordering .}

\begin{figure}
\centering \includegraphics*[width=0.9\linewidth]{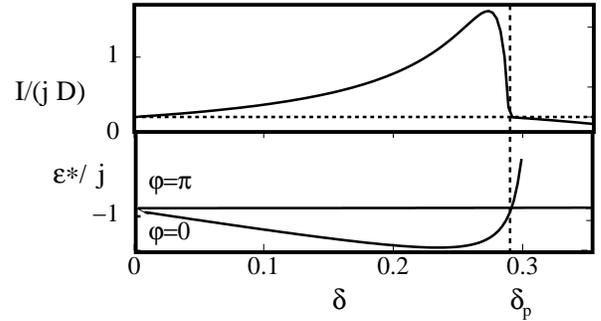}
\caption{Top panel: saturation current $I/(jD)$ vs. $\delta$. For small values 
of $\delta$, $I/(jD)$ 
increases linearly with $\delta$. At large enough $\delta$ it decreases and
becomes 0 at $\delta = \delta_p$, where  the minimum of the energy moves 
to $\varphi=\pi$. For $\delta > \delta_p$ $I<0$, that is, the system has
crossed over to a $\pi$-junction behavior. 
Bottom panel: $\epsilon^*/j$ vs. $\delta$ for $\varphi=0,\pi$. At $\delta =
\delta_p$ the two curves cross and the minimum moves to $\varphi=\pi$.}
\label{figura3}
\end{figure}

Expansion in powers of $\delta $ is no longer meaningful. The energy gain 
due to superconductivity $G_S \sim -\met N(0) \Delta ^2$ becomes dominant 
with respect  to the Kondo energy gain $ \Omega _K ( \delta = 0 ) $.
Kondo correlations  and superconductivity start to compete. The 
minimum  moves from $\varphi = 0$ to $\varphi = \pi$,
as shown in  Fig.\ref{figura3} (bottom panel). Indeed, in the bottom panel
of Fig.\ref{figura3} we plot $\epsilon^*/j$ vs. $\delta$ for $\varphi=0$
and $\varphi=\pi$ and $j=8$. At a critical value of $\delta$, the two curves
cross each other, corresponding to crossing of the minimum energy from
$\varphi=0$ to $\varphi = \pi$.

There is a third regime, not   covered by our approach:
 
{\sl $c) \:\: \Delta >  T_K $ :Superconducting order is dominant.}

 Here the Kondo  interaction can be treated perturbatively. This regime was 
studied  in Ref.\cite{clerk} using a $NCA $ perturbative approach. 
The results in \cite{clerk} show  a change  of the sign of the Josephson 
current that is consistent with our results, as we discuss below.  

The Josephson current is given by:

\[
I_J = -  2e \partial \epsilon ^* [j,\delta,\varphi ]/\partial \varphi 
\]
\noindent

In regime $ a)$, $I_J$ is linear in $j \delta $:

\beq
I_J  =
- 2 e \Omega_K  B \delta  \sin \varphi \;\; .
\label{eq9}
\eneq
\noindent

Therefore, in this regime Kondo correlations strongly increase the 
Josephson current. The sign of the current is the conventional one  
for  Josephson systems, that is, the device behaves as a diamagnetic 
junction. As $\delta$ slightly increases, the dependence of $I_J$ 
on $\varphi$ develops non sinusoidal terms, in a way similar to the result 
in  Ref.\cite{matveev}.

As $\delta$ increases, we enter a new regime in which superconductivity 
competes with Kondo correlations. In this region the Josephson current 
decreases as $\delta$ increases (see top panel of Fig.\ref{figura3}).
As $\delta = \delta_{p}$, the energy minimum moves from $\varphi=0$ to 
$\varphi = \pi$. Correspondingly, although the current is now linear in
$\delta$, its sign is reversed, that is, the  device now behaves as  a
$\pi$-Josephson junction.
These results appear to extend the analysis performed in
Ref.\cite{clerk} in the regime $\Delta > T_K $ to the complementary regime,
$\Delta < T_K$. 

We now provide the physical interpretation of our results in terms of Andreev
bound states in the normal region. In the noninteracting limit,  
the superconducting  gap at the Fermi level $\mu$ in an SNS  sandwich 
generates two localized Andreev states in the normal region 
with energy within the gap.  The one of the two states with  positive
(negative) energy is mostly particle (hole)-like. In this case, Josephson
conduction across the normal region involves both Andreev states.

As interaction is turned on ($J > 0$) and $\Delta \ll T_K$,  pair 
tunneling takes place through the Kondo resonance at  $\mu$,   
whose width ($\sim T_K$) is quite large. At the same time, dot's spin
``disappears'', because of full compensation at the strongly coupled
fixed point \cite{nozieres}. Full screening  does not allow for extra 
occupancy of the dot. Accordingly, we expect  that the  hole-like state 
will be just above $\mu$ but the particle-like state will be pushed quite 
high in energy (the higher is the effective coupling, the higher will 
be the energy of the particle-like state).

    When superconductivity starts to weaken Kondo correlations, 
( regime $b)$ ), the two states are expected to have comparable 
energy, until, at some point, they cross. Such a crossing, where
the particle state energy moves below the hole state energy and the
particle state becomes partially occupied, makes the minimum move from
$\varphi=0$ to $\varphi=\pi$. By increasing $\Delta $ even further, we reach
the regime discussed in \cite{clerk}, where the particle state has 
moved below $\mu$ and the minimum of the energy has moved to $\varphi = \pi $
(see bottom panel of Fig.\ref{figura3}). This is the perturbative regime, 
where the Josephson current takes an overall - sign and the system behaves
as a $\pi$-junction \cite{clerk}.

In conclusion, we have generalized the variational approach introduced in
ref.\cite{nos} to study Josephson conduction through a  quantum dot at 
Coulomb blockade connected to two superconducting leads. Within such an 
approach, we analyzed the region of parameters where the dot lays in the 
strongly coupled Kondo regime and showed that, in this regime, it behaves 
as a diamagnetic junction, with a corresponding ``Kondo enhancement'' in the 
current. Our results appear to be in agreement  with  earlier predictions on 
related systems \cite{matveev}. Moreover, our formalism shows that the 
transition to $\pi$-junction regime takes place much before antiferromagnetic
correlations at the dot can be treated perturbatively, that is,
much before Kondo effect has been disrupted by superconductivity. Finally, 
our  results lead us to predict an enhancement in the saturation current 
in the strongly coupled regime, which should be experimentally detectable 
as a signature of Kondo effect in the quantum dot between two superconductors.

\end{document}